\documentclass[aps,prl,reprint,showpacs,superscriptaddress]{revtex4-1}

\usepackage{graphicx}
\usepackage{subfigure}
\usepackage{color}
\usepackage{amsmath}
\usepackage{amssymb}
\usepackage{epstopdf}
\usepackage[linktocpage,colorlinks=true,linkcolor=blue,citecolor=blue,breaklinks=true]{hyperref}
\usepackage{verbatim}
\usepackage{breakcites}
\usepackage[version=3]{mhchem}

\newcommand{\be}{\begin{equation}}
\newcommand{\ee}{\end{equation}}
\newcommand{\bs}{\boldsymbol}

\begin{document}

\preprint{}

\title{Sea ice motion as a stochastic process}

\author{S. Toppaladoddi}
\affiliation{School of Mathematics, University of Leeds, Leeds LS2 9JT, U.K.}
\affiliation{All Souls College, Oxford OX1 4AL, U.K.}

\email[]{S.Toppaladoddi@leeds.ac.uk}

\date{\today}

\begin{abstract}
We use tools from statistical physics to develop a stochastic theory for the drift of a single Arctic sea-ice floe. Floe-floe interactions are modelled using a Coulomb friction term, with any change in the thickness or the size of the ice floe due to phase change and/or mechanical deformation being neglected. We obtain a Langevin equation for the fluctuating velocity and the corresponding Fokker-Planck equation for its probability density function (PDF). For values of ice compactness close to unity, the stationary PDFs for the individual components of the fluctuating velocity are found to be the Laplace distribution, in agreement with observations. A possible way of obtaining a more general model that accounts for thermal growth and mechanical deformation is also discussed.

\end{abstract}

\maketitle

\section{Introduction} 

The importance of the Arctic ice cover in the climate system stems primarily from its influence on the Earth's radiation budget, which it exerts through its relatively large albedo \citep{OneWatt}. Despite its importance, accurately predicting the spatio-temporal evolution of the ice cover, subject to prescribed forcings, still remains challenging. One of the principal challenges associated with making this prediction is the dynamics of the ice cover \citep{OneWatt, Rothrock:1975, rampal2011}. 

The ice cover is not continuous, but is made up of a very large number of floes of different shapes, sizes, and thicknesses \citep{Thorndike1975, Rothrock:1984}. It can be inferred from field and satellite observations that sea ice behaves differently at different length scales: At the scale of an individual floe it moves like a deformable solid body, but at basin-wide scales it moves like a highly viscous liquid. This suggests that the following two approaches can be used to study its motion (see Solomon \citep{solomon1970} for a more general discussion): 
\begin{enumerate}
\item In the first approach, the motion of individual ice floes is studied by solving Newton's equations for each floe. From these solutions, one then extracts statistical information \citep{rampal2009, agarwal2017} that is used to describe the motion of the ice cover at much larger length scales. 
\item And in the second, one takes the ice cover to be a continuum and develops rheological models to relate the internal stress field to the other macroscopic variables, including the thickness distribution of ice \citep{Thorndike1975, TW2015, TW2017, TMW2023}. This is then used in the Cauchy equation, with appropriate boundary conditions, to solve for the velocity field. 
\end{enumerate}
A principal aim of both these approaches is to predict the statistical properties of the ice velocity field \citep{thorndike1982, colony1984, colony1985, thorndike1986, rampal2009, agarwal2017}. Both approaches have their advantages and limitations, but the focus since the Arctic Ice Dynamics Joint Experiment (AIDJEX) expedition has been on developing observationally consistent rheological models \citep{Rothrock:1975, Feltham:2008}.

The internal stress field in the ice cover is a consequence of the mechanical interactions between the constituent ice floes \citep{Rothrock:1975}. However, unlike in the kinetic theory of gases, where molecules are assumed to interact only via elastic collisions \citep{Harris}, there are different modes of interaction -- rafting, ridging, shearing, and jostling -- between ice floes \citep{VW08}. This makes it more challenging to develop a Boltzmann-like theory for sea ice velocity, although the development of such a theory has been attempted in the context of Saturn's rings \citep{goldreich1978}, where the last three modes of interaction between ice particles are possible. 

The first attempt to include floe-floe interactions into the dynamics of a single ice floe was by Sverdrup \citep{sverdrup1928}. He introduced a frictional force proportional to the floe velocity, but always in the direction opposite to it. However, this formulation is not completely correct as friction can both decelerate and accelerate an object depending on the relative velocity \citep{reed1962, solomon1970}. A more generalized description of the floe-floe interactions was developed by \citet{solomon1970} and \citet{timokhov1970}, who considered deterministic and stochastic drift of ice in one dimension, respectively. The stochasticity in Timokhov's model \citep{timokhov1970} was introduced through the probability for dynamic coagulation to occur when ice floes collided. A drawback of both these models is in the introduction of spatial gradients (of velocity in Solomon's model and of compactness in Timokhov's), which makes it unclear at what length scales the continuum assumption holds in these models. In more recent work, the ice cover has been treated as a two-dimensional granular gas to study the emergence of the internal stress from floe-floe collisions \citep{shen1987}, and flow \citep{feltham2005} and clustering \citep{herman2011} of ice floes in marginal ice zones.

The collective motion of ice floes on length scales much larger than the floe size can be approximated as that of a continuum \citep{Thorndike1975}. In this case, the equations that describe the evolution of the ice cover are the mass balance and Cauchy equations. The principal challenge associated with this approach has been in determining the constitutive equation for the internal stress \citep{Rothrock:1975}. Observations made during the AIDJEX project motivated the development of the elastic-plastic rheological model of sea ice \citep{Coon1974}. Subsequently, other rheological models have been proposed to capture various features observed in pack ice \citep{wilchinsky2004, girard2011, tsamados2013, bouillon2015, dansereau2016}. A detailed discussion of the theoretical underpinnings of some of the rheological models can be found in the reviews by \citet{Rothrock:1975} and \citet{Feltham:2008}.

The principal aim of the current work is to develop a stochastic theory of sea ice motion to capture some of the observed statistical properties \citep{rampal2009}. We achieve this by extending Sverdrup's model for the floe-floe interaction by introducing a Coulomb friction term that accounts for both acceleration and deceleration of a single ice floe. Our model is analogous to that of a Brownian particle subjected to viscous and dry frictional forces in an external force field \citep{de2005, hayakawa2005}. Our formulation of the problem permits us to explicitly calculate the probability density functions (PDFs) of the components of the fluctuating velocity, which are then compared with observations \cite{rampal2009}.

\section{The new theory} 

We consider the ice floes to be rigid circular discs with thickness $h$ and radius $R$. Focussing on one of these floes, the governing equations for the horizontal motion are:
\be
\frac{d \bs{x}}{dt} = \bs{v},
\label{eqn:position}
\ee
and
\be
\frac{d}{dt} \left(m \, \bs{v}\right) = \bs{F_a} + b \, \bs{\xi}(t) + \bs{F_o} - 2 \, m \, \Omega \, \bs{k} \times \bs{v} - \mathcal{F} \, \bs{S}(\bs{v} - \bs{\left<v\right>}).
\label{eqn:velocity}
\ee
Here, $\bs{x} = (x,y)$ is the position vector, $m$ is the mass of the ice floe, $\bs{v} = (u, v)$ is its two-dimensional velocity, $\bs{F_a}(\bs{x},t)$ is the mean wind force, $b \, \bs{\xi(t)}$ represents the fluctuations in the wind forcing with $b$ being the amplitude of the fluctuations and $\bs{\xi(t)}$ being Gaussian white noise, $\bs{F_o}(\bs{x},t)$ represents the ocean drag force, $\Omega$ is the Coriolis frequency and $\bs{k}$ is the unit vector along the vertical, $\mathcal{F}(C)$ is the threshold value of the Coulomb friction force due to the neighbouring ice floes, $C \left(\in \left[0, 1\right]\right)$ is the constant compactness of the ice cover, and $\bs{S}$ is a vector given by
\be
\bs{S}(\bs{v} - \bs{\left<v\right>}) = \frac{\bs{v} - \left<\bs{v}\right>}{|\bs{v} - \left<\bs{v}\right>|},
\label{eqn:sign1}
\ee 
where $\left<...\right>$ denotes the ensemble average. The function $\bs{S}$ is a generalization of the sign function for a two-dimensional vector. Introducing the fluctuating velocity as $\bs{v'} = \bs{v} - \bs{\left<v\right>}$, we can write equation \ref{eqn:sign1} as
\be
\bs{S}(\bs{v'}) = \frac{\bs{v'}}{|\bs{v'}|}.
\ee
For simplicity, we have neglected the forces due to horizontal pressure gradient of the atmosphere and gradients in sea surface height; but, they can be included in the model without any difficulty.

In introducing the floe-floe interaction term in equation \ref{eqn:velocity}, we have made the following assumptions: (a) interactions that only involve pushing and/or shearing between ice floes are important; (b) the role of collisions here is to drive the velocity to its mean value, $\left<\boldsymbol{v}\right>$; and (c) the value of the threshold force varies linearly with compactness, i.e., $\mathcal{F}(C) = \mathcal{F}_0 \, C$, where $\mathcal{F}_0$ is a constant. 

The reasoning for this model is the following. If $N (\gg 1)$ is the total number of ice floes, then any description of a single ice floe requires us to take into account its interactions with its $n (\ll N)$ nearest neighbours. However, the construction of a system of coupled deterministic/stochastic differential equations for these localized interactions would also require us to take into account the interactions of the neighbouring ice floes with their own nearest neighbours. Hence, any description of the dynamics of a subset of the $N$ ice floes leads to a closure problem, which is similar to the closure problem encountered in the kinetic theory of gases \citep{Harris}. Consequently, some assumptions would have to be made to truncate the problem. The mathematical form of the interactions in equation \ref{eqn:velocity}, along with assumption (b) above, represents a mean-field approximation of these interactions. This mean-field interaction term embodies the fact that the collisions between the ice floe and its neighbours are due to the different velocities with which they move. However, if they all were to move with the average velocity $\left<\boldsymbol{v}\right>$ then they would not collide with each other, in which case $\bs{S} = \bs{0}$. We should also note here that the mathematical form of the floe-floe interactions in equation \ref{eqn:velocity} permits both acceleration and deceleration of the ice floe depending on the sign of the fluctuation. Furthermore, assuming the ocean drag is proportional to the ice velocity \citep{lepparanta2011} and the ocean is at rest, equation \ref{eqn:velocity} becomes
\be
\frac{d \bs{v}}{dt} = \bs{F_a} + b \, \bs{\xi}(t) - \beta \, \bs{v} - 2 \, \Omega \, \bs{k} \times \bs{v} - \mathcal{F} \, \bs{S}(\bs{v'}),
\label{eqn:velocity2}
\ee
where $m$ has been set to unity without any loss of generality and $\beta$ is a constant. 

To obtain an equation for the velocity fluctuations, we first take the mean of equation \ref{eqn:velocity2} giving
\be
\frac{d \left<\bs{v}\right>}{dt} = \bs{F_a} - \beta \, \left<\bs{v}\right> - 2 \, \Omega \, \bs{k} \times \left<\bs{v}\right>.
\label{eqn:velocity3}
\ee
It is seen that the mean velocity of the ice floe is unaffected by the collisions with other ice floes. This is due to the fact that these collisions lead to both accelerating and decelerating forces and an average over the ensemble gives zero net force. Subtracting equation \ref{eqn:velocity3} from equation \ref{eqn:velocity2} and neglecting the effect of Coriolis force on fluctuations, we get
\be
\frac{d \bs{v'}}{dt} = - \beta \, \bs{v'} - \mathcal{F} \, \bs{S}(\bs{v'}) + b  \, \bs{\xi}(t),
\label{eqn:fluctuations}
\ee
which is the required equation. The corresponding generalized Fokker-Planck equation -- also called the Kramers-Chandrasekhar equation -- is given by
\be
\frac{\partial P}{\partial t} + \left(\left<\bs{v}\right> + \bs{v'}\right) \cdot \nabla P = \nabla_{\bs{v'}} \cdot \{\left[\beta \, \bs{v'} + \mathcal{F} \, \bs{S}(\bs{v'})\right] \, P + D \, \nabla_{\bs{v'}} P\}.
\ee
Here, $P \equiv P(\bs{x}, \bs{v'},t)$ is the PDF for the velocity fluctuations and $D = b^2/2$. Assuming $P$ is spatially homogeneous leads to
\be
\frac{\partial P}{\partial t} = \nabla_{\bs{v'}} \cdot \{\left[\beta \, \bs{v'} + \mathcal{F} \, \bs{S}(\bs{v'})\right] \, P + D \, \nabla_{\bs{v'}} P\},
\label{eqn:FPE}
\ee
which is the required evolution equation for the PDF of the velocity fluctuations.

\section{Results}

\subsection{Stationary solution}
To obtain the stationary solution to equation \ref{eqn:FPE}, we introduce the drift vector $\bs{\mathcal{D}}$, which is defined as
\be
\bs{\mathcal{D}} \equiv - \left[\beta \, \bs{v'} + \mathcal{F} \, \bs{S}(\bs{v'})\right] = -\left(\beta \, \bs{v'} + \mathcal{F} \, \frac{\bs{v'}}{|\bs{v'}|}\right).
\ee
In component form, this is written as
\be
\left(\mathcal{D}_{u'}, \mathcal{D}_{v'}\right) = - \left(\beta \, u' + \mathcal{F} \, \frac{u'}{\sqrt{u'^2 + v'^2}}, \beta \, v' + \mathcal{F} \, \frac{v'}{\sqrt{u'^2 + v'^2}}\right).
\label{eqn:Dcomp}
\ee
Now, it is easily seen from equation \ref{eqn:Dcomp} that
\be
\frac{\partial \mathcal{D}_{u'}}{\partial v'} = \frac{\partial \mathcal{D}_{v'}}{\partial u'} = \mathcal{F} \, \frac{u' \, v'}{\left(u'^2+v'^2\right)^{3/2}},
\ee
which implies that $\bs{\mathcal{D}}$ can be expressed as the gradient of a potential, i.e., $\bs{\mathcal{D}} = - \nabla_{\bs{v'}} \Phi$, and that $\bs{\mathcal{D}}$ vanishes at $u' = v' = \pm \infty$ in the stationary state \citep{risken1996}. The stationary solution is readily found to be \citep[see e.g., chapter 6 in][]{risken1996}
\be
P(u',v') = \mathcal{N} \, \exp{\left(-\frac{1}{D} \, \Phi\right)},
\ee
where $\mathcal{N}$ is the integration constant determined by requiring that $P$ is normalized, and 
\be
\Phi = - \left(\int \mathcal{D}_{u'} \, du' + \int \mathcal{D}_{v'} \, dv'\right).
\ee
Using equation \ref{eqn:Dcomp} to evaluate the integrals, we obtain
\be
\Phi = \frac{\beta}{2} \, (u'^2 + v'^2) + 2 \, \mathcal{F} \, \sqrt{u'^2 + v'^2},
\ee
and hence
\be
P(u',v') = \mathcal{N} \, \exp{\left(-\frac{1}{D} \, \left[\frac{\beta}{2} \, (u'^2 + v'^2) + 2 \, \mathcal{F} \, \sqrt{u'^2 + v'^2}\right]\right)},
\label{eqn:FPE_ss}
\ee
which is the required stationary solution for the most general case. In the following, we consider two different regimes of sea ice drift based on the values of the compactness.

\subsection{Regime 1: Low value of compactness ($C\approx 0$)}

For very low values of $C$, the effects of the neighbouring ice floes is negligible. Hence, in this regime, $\mathcal{F} = 0$ and the normalized stationary solution is
\be
P(u',v') = \frac{\beta}{2 \, \pi \, D} \, \exp{\left(-\frac{\beta}{2 \, D} \, \left(u'^2 + v'^2\right)\right)},
\label{eqn:vel_Gauss}
\ee
which is the well-known solution to the classical Brownian motion problem \citep{chandra1943}. The fluctuations in the ice-floe velocity are due to the fluctuating wind, which has been assumed to be Gaussian in nature. This leads to the ice-floe velocity fluctuations being Gaussian as well.

\subsection{Regime 2: High value of compactness ($C\approx 1$)}

In this regime, the ice floe undergoes continuous collisions with the neighbouring ice floes. Hence, it is natural to assume here that most of the resistance to the random motion of the ice floe comes from its neighbours. Setting the ocean drag force to zero ($\beta = 0$), we get the normalized stationary solution in this regime to be
\be
P(u',v') = \frac{2 \, \mathcal{F}_0^2}{\pi \, D^2} \, \exp{\left(-\frac{2 \, \mathcal{F}_0}{D} \, \sqrt{u'^2 + v'^2}\right)}.
\label{eqn:2D_coulomb}
\ee
The PDFs for the individual components can be found using
\be
P_{u'} = \int_{-\infty}^{\infty} P(u',v') \, dv' =  \int_{-\infty}^{\infty} \frac{\Lambda^2}{2 \, \pi} \, \exp{\left(-\Lambda \, \sqrt{u'^2 + v'^2}\right)} \, dv',
\label{eqn:single_PDF_u}
\ee
and similarly
\be
P_{v'} = \int_{-\infty}^{\infty} P(u',v') \, du' =  \int_{-\infty}^{\infty} \frac{\Lambda^2}{2 \, \pi} \, \exp{\left(-\Lambda \, \sqrt{u'^2 + v'^2}\right)} \, du',
\label{eqn:single_PDF_v}
\ee
where  $\Lambda = \frac{2 \, \mathcal{F}_0}{D}$. We should note here that $P_{u'}$ and $P_{v'}$ have identical functional forms. The integrals in equations \ref{eqn:single_PDF_u} and \ref{eqn:single_PDF_v} cannot be evaluated to give analytical expressions for $P_{u'}$ and $P_{v'}$, so they would have to be computed numerically.

The components $u'$ and $v'$ enter equation \ref{eqn:2D_coulomb} as $\sqrt{u'^2 + v'^2}$. This makes obtaining the PDF for the fluctuating speed, $\mathcal{P}(V')$, straightforward, and is readily found to be \citep[see e.g., chapter 7 in][]{Reif}
\be
\mathcal{P}(V') = \Lambda^2 \, V' \, \exp{\left(- \Lambda \, V'\right)}, 
\label{eqn:speed_PDF}
\ee
where $V' = \sqrt{u'^2 + v'^2}$.

\subsection{Comparison with observations}
To compare the PDFs from our theory with observations, we use the results from the analysis of the International Arctic Buoy Program data from \citet{rampal2009}. A total of 450 drifters deployed between 1979 and 2001 were used, and data from only those buoys whose positions were at least 100 kms away from the coasts were chosen. \citet{rampal2009} chose a two-dimensional Cartesian co-ordinate system centered at the North Pole, with one of the axes pointing along the Greenwich meridian. For the analysis, the time period for winter was chosen from November to mid May, and for summer from mid June to mid September. Further details on the procedure used to obtain the mean velocity, including the choice of length and time scales for averaging, and the velocity fluctuations can be found in \citet{rampal2009}.

\begin{figure}
\centering
\includegraphics[trim = 100 0 150 0, scale=0.18]{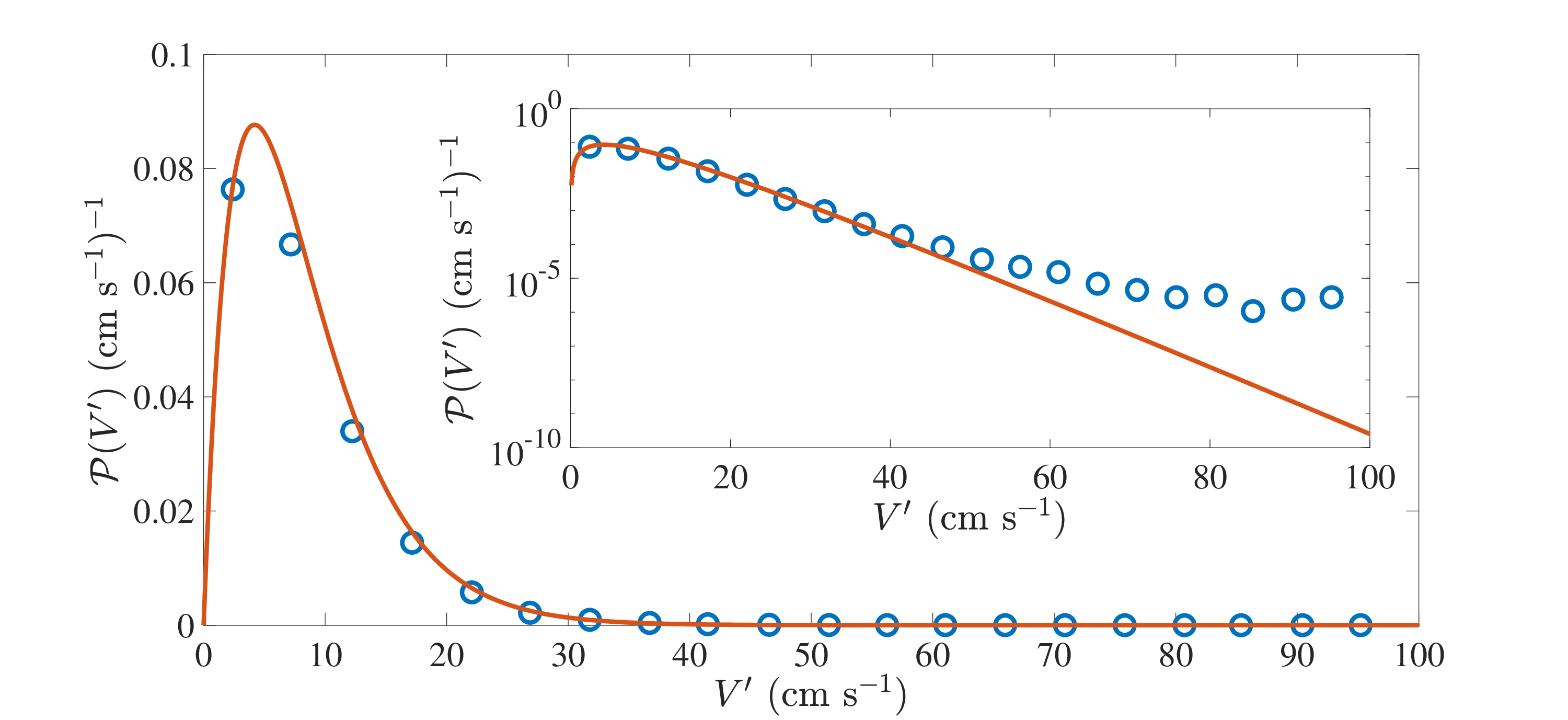}
\caption{Comparison of our theoretical PDF for the fluctuating speed with observations. Circles are data from \citet{rampal2009} and the solid curve is the functional form of the solution from theory (equation \ref{eqn:speed_PDF}). The value of $\Lambda$ obtained from the fit is $0.238$ (cm/s)$^{-1}$. The inset shows the same figure in log-linear plot.}
\vspace{-5mm}
\label{fig:speed_pdf}
\end{figure}

In figure \ref{fig:speed_pdf} we show the comparison between the theoretical PDF for the fluctuating speed and the observational PDF from Rampal \emph{et al.} \citep{rampal2009}. The functional form of the solution (equation \ref{eqn:speed_PDF}) is fit to the observational data, and $\Lambda$, which is the only fitting parameter, is determined from this fit.

\begin{figure}
\centering
\includegraphics[trim = 100 0 150 0, scale=0.18]{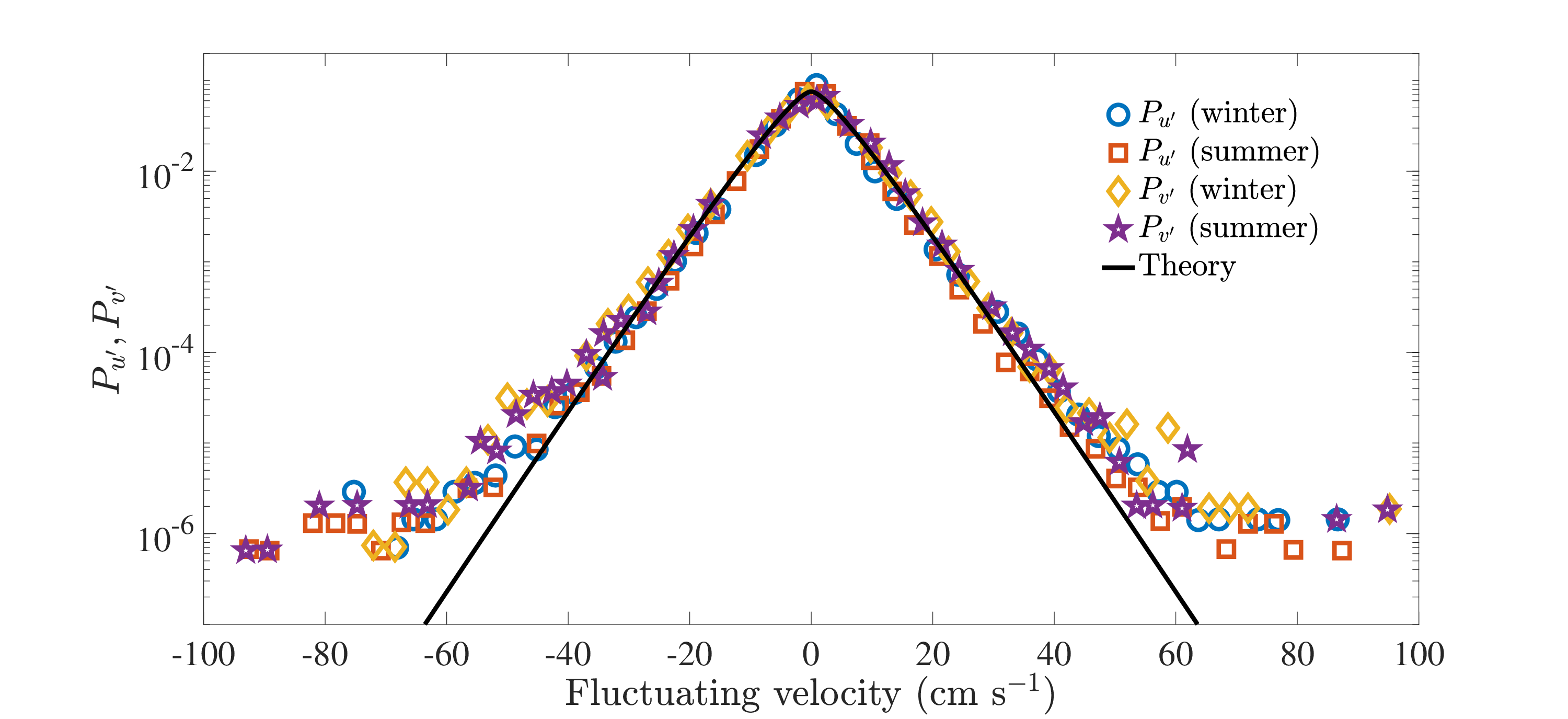}
\caption{Comparison of our theoretical PDFs for sea-ice velocity fluctuations with observations. Symbols are data from \citet{rampal2009} and the solid curve is the PDF obtained from equation \ref{eqn:single_PDF_u} after numerically evaluating the integral. The value of $\Lambda$ used is $0.238$ (cm/s)$^{-1}$ (see figure \ref{fig:speed_pdf}).}
\vspace{-5mm}
\label{fig:velocity_pdf}
\end{figure}

Using the value of $\Lambda = 0.238$ (cm/s)$^{-1}$ from the fit in figure \ref{fig:speed_pdf}, we compute the integral in equation \ref{eqn:single_PDF_u} numerically to determine $P_{u'}$. Noting that $P_{v'}$ and $P_{u'}$ have the same functional forms (equations \ref{eqn:single_PDF_u} and \ref{eqn:single_PDF_v}), we compare the PDF with observations and find that the PDFs for the velocity components are Laplace distributions. This is shown in figure \ref{fig:velocity_pdf}. As noted by \citet{rampal2009}, it is remarkable that the PDFs for $u'$ and $v'$ are approximately the same in each season, showing the fluctuations are isotropic. 

It is interesting to note that the Gaussian distribution (equation \ref{eqn:vel_Gauss}) is not observed for the period between 1979-2001. However, with the dramatic decline in the ice cover over the last two decades, it might now be possible to test the correctness of this prediction. This is a part of our ongoing work.

\section{Conclusions}

We have developed a stochastic theory for the drift of a single ice floe in the Arctic. The floe-floe interactions are introduced through the Coulomb friction term \citep{de2005, hayakawa2005} in the equation of motion. We first obtained the Langevin equation for the velocity fluctuations, and then the corresponding Fokker-Planck equation. We found that for values of compactness close to unity, the stationary PDFs of the individual fluctuating velocity components are the Laplace distribution. However, for very small values of compactness, we obtained Gaussian distribution as the stationary solution. Comparison of the functional form of solution for $C \approx 1$ with observations \citep{rampal2009} shows good qualitative agreement. This agreement, despite the many simplifying assumptions made, provides confidence that the mathematical formulation of the problem is physically sound, and that the model captures the leading order physics associated with the sea-ice velocity fluctuations.

However, a shortcoming of the current model is that it does not take into account the thermal growth and mechanical deformations of the ice floe. This is remedied by writing the mass of the ice floe as $m = \rho_i \, \pi \, R^2 \, h$, where $\rho_i$ is the constant density of the ice floe, and rewriting equation \ref{eqn:velocity} as
\be
m \frac{d \bs{v}}{dt} = - \frac{d m}{dt} \, \bs{v} + \bs{F_a} + b \, \bs{\xi}(t) + \bs{F_o} - m \, \Omega \, \bs{k} \times \bs{v} - \mathcal{F} \, \bs{S}(\bs{v} - \bs{\left<v\right>}).
\label{eqn:coupled}
\ee
The changes in $R(t)$ and $h(t)$ can now be coupled with the momentum equation (equation \ref{eqn:coupled}). This, however, leads to a six-dimensional Fokker-Planck equation which is very challenging to solve -- both analytically and numerically. In such a situation, it might be more prudent to solve the coupled stochastic differential equations for $\bs{v}$, $h$ and $R$. The previous work on the thickness distribution of sea ice \citep{TW2015, TW2017, TMW2023} and our current theory provide a physical and mathematical framework to explore these coupled problems in future.  

\section*{Acknowledgements}
The author thanks A. J. Wells for his critical comments on a previous version of the manuscript, which were helpful in improving this work.

\bibliography{g(h)}

\end{document}